\RequirePackage{amsmath}

\documentclass[12pt]{iopart}
\pdfoutput=1
\usepackage[caption=false,captionskip=-0.5mm]{subfig}
\usepackage{graphicx}
\usepackage{amsmath}
\usepackage{amsfonts}
\usepackage{mathtools}
\usepackage{amssymb}
\usepackage{xcolor}

\colorlet{mylinkcolor}{blue!66!black!80}

\usepackage[colorlinks=true,linkcolor=mylinkcolor,citecolor=mylinkcolor,filecolor=cyan,urlcolor=mylinkcolor,breaklinks=true]{hyperref}

\usepackage[utf8]{inputenc}

\usepackage{cite}

\DeclareMathAlphabet{\mathcal}{OMS}{cmsy}{m}{n}
\newcommand{\avg}[1]{\langle#1\rangle}

\newcommand{\del}{\partial}
\newcommand{\dd}{{\rm d}}
\newcommand{\ee}{{\rm e}}

\begin{document}
\title{Nonequilibrium sensing and its analogy to kinetic proofreading}
\author{David Hartich, Andre C. Barato, and Udo Seifert}
\address{II. Institut f\"ur Theoretische Physik, Universit\"at Stuttgart, 70550 Stuttgart, Germany}
\eads{\href{mailto:hartich@theo2.physik.uni-stuttgart.de}{hartich@theo2.physik.uni-stuttgart.de}, \href{mailto:barato@theo2.physik.uni-stuttgart.de}{barato@theo2.physik.uni-stuttgart.de} and
\href{mailto:useifert@theo2.physik.uni-stuttgart.de}{useifert@theo2.physik.uni-stuttgart.de}}
\date{\today}

\begin{abstract}
For a paradigmatic model of chemotaxis, we analyze the effect of how a nonzero affinity driving receptors out of equilibrium affects 
sensitivity. This affinity arises whenever changes in receptor activity involve ATP hydrolysis. The 
sensitivity integrated over a ligand concentration range is shown to be enhanced by the affinity, providing 
a measure of how much energy consumption improves sensing. With this integrated sensitivity we can establish
an intriguing analogy between sensing with nonequilibrium receptors and kinetic proofreading: the increase in integrated sensitivity
is equivalent to the decrease of the error in kinetic proofreading. The influence 
of the occupancy of the receptor on the phosphorylation and dephosphorylation reaction rates is shown to be crucial for the relation between 
integrated sensitivity and affinity. This influence can even lead to a regime where a nonzero affinity decreases the integrated sensitivity,
which corresponds to anti-proofreading.
\end{abstract}

\section{Introduction}
Bacterial chemotaxis, a process by which the cell directs its motion
in response to external ligand concentration, is a canonical example
of biological sensing. Experiments with {\sl E. coli} have provided much insight
into chemotaxis \cite{berg72,berg04}, making this bacterium sensory system a particularly well
understood example. In {\sl E. coli} chemotaxis, the sensitivity is a
key observable quantifying the response in activity inside the cell due to a change
in the external ligand concentration.     

Stochastic models for {\sl E. coli} \cite{tu08,mell07,tu13,clau14} receptors often assume
that changes in activity are described by an equilibrium process
involving only conformational changes, leading to 
an equilibrium Monod-Wyman-Changeux (MWC) model  \cite{mono65,marz13}. However,
chemical reactions where the receptor changes from an inactive
to an active state often involve free energy consumption through, for example, adenosine triphosphate (ATP) hydrolysis. A stochastic
model including this feature must have transition rates that break
detailed balance leading to a nonzero affinity, corresponding to the chemical potential difference involved in ATP hydrolyisis,
driving the process out of equilibrium.
 
Two recent studies have analyzed the effect of such an affinity
in models related to the {\sl E. coli} sensory network. Tu \cite{tu08a} has considered
the effect of the driving affinity on both the dwell-time distribution and the
sensitivity in a model for the flagellar motor switching between run and tumble.
Skoge et al. \cite{skog13} have shown that nonequilibrium receptors can increase
the signal to noise ratio for fixed sensitivity.

Beyond {\sl E. coli} chemotaxis, the effect of energy dissipation
in biological processes involving information processing has received much attention recently \cite{qian05,meht12,lan12,palo13,lang14,gove14,gove14a,sart14,bara13a,bara14b,bo15}.
A prominent example among such processes is kinetic proofreading \cite{hopf74,nini75,benn79,ehre80}, which is a dissipative error reduction 
mechanism related to copying biochemical information. As this error reduction is achieved through free energy consumption, a nonzero affinity
driving the process out of equilibrium is also present in kinetic proofreading. Specifically, relations between the error and the driving affinity
have been obtained \cite{ehre80,qian06} (see also  \cite{qian07,qian08,qian10,muru12,sart13,muru14} for other recent works).   
 
In this paper, we consider a nonequilibrium model for {\sl E. coli} receptors including ATP hydrolysis in the chemical reactions that involve changes in activity.
We quantify the effect of having a nonzero driving affinity on sensing by analyzing an integrated sensitivity, which is an integral of the sensitivity over a concentration range.
This observable is shown to have a simple relation with the affinity driving the process out of equilibrium. We show that sensing with 
nonequilibrium receptors and kinetic proofreading can be viewed as equivalent problems, with the increase in the integrated sensitivity in nonequilibrium sensing
being analogous to the error reduction in kinetic proofreading. 

The transition rates for changes in activity are assumed to depend on whether the receptor is occupied by a ligand or empty. We show that this dependency is
quite important for the relation between sensitivity and the driving affinity. There is even a regime where energy dissipation leads to a decrease in the integrated sensitivity, 
which is equivalent to an anti-proofreading regime in kinetic proofreading \cite{muru14}.

The paper is organized as follows. Section \ref{sec:Main_result} contains a simple stochastic model for a single  nonequilibrium receptor. In section \ref{sec3} we introduce
the integrated sensitivity and obtain its relation with the affinity driving the process out of equilibrium. The analogy between nonequilibrium sensing and
proofreading is established in section \ref{sec4}. In section \ref{sec:multiple_links}, with a more general model for a single receptor, we analyze how the influence of the occupancy of the receptor on the
phosphorylation and dephosphorylation reaction rates affects the relation between integrated sensitivity and affinity. We conclude in section \ref{sec:conclusions}. 
Moreover,  \ref{sec:Nsites} contains a generalization of the single receptor model analyzed in the main text to an arbitrary number of binding sites.

\section{Nonequilibrium receptor model}
\label{sec:Main_result}%
\begin{figure}%
 \centering%
  \includegraphics{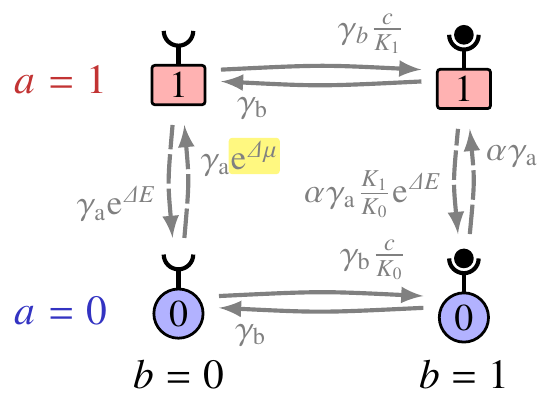}
 \caption{Four-state model for a single receptor. Vertical transitions correspond to a change in activity, while horizontal transitions correspond to a change in the occupancy of the receptor. 
The phosphorylation rates in \eqref{eq:chem_phos} are chosen as $\kappa_+^0=\gamma_\mathrm{b}\ee^{\varDelta\mu}$ and $\kappa_-^0=\gamma_\mathrm{a}\ee^{\varDelta E}$. The dephosphorylation rates in \eqref{eq:chem_dephos} are chosen as 
$\omega_+=\alpha\gamma_\mathrm{a}(K_1/K_0)\ee^{\varDelta E}$ and $\omega_-=\alpha\gamma_\mathrm{a}$.}%
 \label{fig:general4state}%
\end{figure}

The single receptor model we analyze in this paper is defined as follows (see  Fig. \ref{fig:general4state}).
There are two binary variables $a$ and $b$ characterizing the state of the receptor, with $b=1$ if the receptor is occupied by a ligand (bound) and $b=0$ if the receptor is free (unbound), and $a=1$ if
the receptor is active and $a=0$ if the receptor is inactive.

In equilibrium the free energy of the four different states can be written as \cite{marz13,bara14b} 
\begin{equation}
F(a,b)\equiv a\varDelta E-b\ln\frac{c}{K_a},
\label{eq:Fab}
\end{equation}
where $\varDelta E$ is the conformational energy difference between active and inactive for a free receptor ($b=0$), $K_a$ is the dissociation constant that
depends on the activity $a$, and $c$ is the external ligand concentration. Setting Boltzmann's constant and the temperature to $k_{\rm B}T\equiv1$, the equilibrium stationary probability is $P_{a,b}\propto\exp[-F(a,b)]$.
Denoting the coarse-grained probability by $P_a\equiv \sum_{b'} P_{a,b'}$, we obtain
\begin{equation}
\left.\frac{P_0}{P_1}\right|_\mathrm{eq}
=\ee^{\varDelta E}\left(\frac{1+\frac{c}{K_0}}{1+\frac{c}{K_1}}\right),
\label{eq:fraction_special_DB}
\end{equation}
where $P_0=1-P_1$. The average activity
\begin{equation}
\avg{a}_c\equiv \sum_{a}aP_{a}
\end{equation}
is just $\avg{a}_c=P_1$. It is assumed that the dependence of the dissociation constant on the 
activity is such that $K_1> K_0$, i.e., the free energy barrier for binding a ligand to an inactive receptor is smaller.  
Hence, from Eq. \eqref{eq:fraction_special_DB} the average activity is a decreasing function of the concentration. This single receptor MWC model already contains the key feature of self-regulation.
However, in order to have cooperativity, which is another important feature of the MWC model, we need more than one binding site \cite{marz13}. The generalization of the model
to an arbitrary number of binding sites is contained in \ref{sec:Nsites}. For our present purposes it is more convenient to restrict to a single binding site in the main text.  

We now consider a nonequilibrium model that includes ATP hydrolysis. For simplicity we assume that when the receptor is unbound only phosphorylation takes place and when the receptor is
bound only dephosphorylation occurs. A more general model with phosphorylation and dephosphorylation occurring for both bound and unbound states, and the implications of 
this generalization are discussed in section \ref{sec:multiple_links}.
The phosphorylation reaction is represented as 
\begin{equation}
 \mathrm{ATP}+
 \raisebox{-0.2\height}{\includegraphics{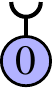}}
  \xrightleftharpoons[\kappa^0_-]{\kappa^0_+}
\mathrm{ADP}+
\raisebox{-0.2\height}{\includegraphics{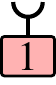}}
,	\label{eq:chem_phos}
\end{equation}
where $\kappa_+^0$ and $\kappa_-^0$ denote transition rates. The dephosphorylation reaction reads  
\begin{equation}
\raisebox{-0.2\height}{\includegraphics{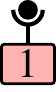}}
  \xrightleftharpoons[\omega^1_-]{\omega^1_+}
  \raisebox{-0.2\height}{\includegraphics{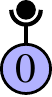}}
  +\mathrm{P_i},
	\label{eq:chem_dephos}
\end{equation}
where $\omega_+^1$ and $\omega_-^1$ are transition rates. 

With the free energy \eqref{eq:Fab}, the generalized detailed balance relation \cite{seif12} imposes 
the following constraints on the rates. First, we have
\begin{equation}
\ln\frac{\kappa^0_+\omega^1_+}{\kappa^0_-\omega^1_-}= \varDelta\mu+\ln\frac{K_1}{K_0},
\label{eq:dmuDef}
\end{equation}
where $\varDelta \mu= \mu_{\rm ATP}-\mu_{\rm ADP}-\mu_{\rm P_i}$ is the free energy dissipated in one ATP hydrolysis.
Second, the transition rates from $b=0$ to $b=1$, denoted by $w_{01}^{a}$, and from $b=1$ to $b=0$, denoted $w_{10}^{a}$, fulfill the relation  
\begin{equation}
\ln\frac{w_{01}^{a}}{w_{10}^{a}}= \ln (c/K_a).
\label{eq:dmuDef2}
\end{equation}
With these two constraints the product of the transition rates in a cycle in the clockwise direction
in Fig. \ref{fig:general4state} is precisely $\varDelta \mu$, which is the affinity driving the process out of equilibrium.
For simplicity we use the specific transition rates given in Fig. \ref{fig:general4state}. The parameters 
$\gamma_{\rm a}$ and $\gamma_{\rm b}$ set the time-scale of the active/inactive and bound/unbound transitions, respectively.
The parameter $\alpha$ is related to redistributing energy weights among the transition rates in such way that the constraints \eqref{eq:dmuDef} and \eqref{eq:dmuDef2} are still fulfilled.

A reasonable assumption is that ligand binding is much faster than activity changes, i.e., $\gamma_\mathrm{a}/\gamma_\mathrm{b}\ll1$ \cite{tu08}.
With this assumption, calculating the stationary probability distribution using standard methods \cite{schn76,hill05} we obtain 
\begin{equation}
\frac{P_0}{P_1}
=\ee^{\varDelta E-\varDelta\mu}\left(\frac{1+\frac{c}{K_0}}{1+\frac{c}{K_1}}\right)\left(\frac{1+\alpha\frac{c}{K_0}}{1+\alpha\frac{c}{K_0}\ee^{-\varDelta\mu}}\right).
\label{eq:fraction_special}
\end{equation}
Comparing with the equilibrium expression \eqref{eq:fraction_special} there is the extra term of the second brackets, which becomes $1$
for $\varDelta\mu=0$. The precise effect of this extra term in sensing is discussed in the next section. For this discussion it is convenient to define the
effective dissociation constants
\begin{equation}
 \tilde{K}_0\equiv \frac{K_0}{\alpha}\quad \text{and}\quad \tilde{K}_1\equiv \frac{K_0}{\alpha}\ee^{\varDelta\mu}.
 \label{eq:Keff}
\end{equation}

\section{Integrated sensitivity}
\label{sec3}
A key observable in sensing is the sensitivity 
\begin{equation}
R(c,\varDelta E)\equiv-4\frac{\del}{\del \ln c}\avg{a}_c=-4\frac{\del P_1}{\del \ln c},
\label{eq:Def_sensitivity}
\end{equation}
which is the response of the average activity to small changes in ligand concentration. 
It is convenient to rewrite the sensitivity as 
\begin{equation}
 R(c,\varDelta E)=4P_0P_1\frac{\del}{\del \ln c}\ln\frac{P_0}{P_1}\le \frac{\del}{\del \ln c}\ln\frac{P_0}{P_1},
 \label{eq:fmax_def}
\end{equation}
where the inequality comes from $P_0P_1\le 1/4$. From Eq. \eqref{eq:fraction_special} it follows that the  upper bound on the right hand side of Eq. \eqref{eq:fmax_def} does not depend on $\varDelta E$.
Hence, for given $c$ there is an optimal conformational free energy difference that maximizes the sensitivity
\begin{equation}
\varDelta E^*(c)=\varDelta\mu+\ln\left(\frac{1+\frac{c}{K_1}}{1+\frac{c}{K_0}}\right)+\ln\left(\frac{1+\alpha\frac{c}{K_0}\ee^{-\varDelta\mu}}{1+\alpha\frac{c}{K_0}}\right),
\label{eq:E*}
\end{equation}
which is obtained from Eq. \eqref{eq:fraction_special} with $P_0P_1=1/4$. 
A free energy close to this optimal value can be achieved through an adaptation system that uses the methylation levels to adjust  $\varDelta E$ in accordance with the external concentration \cite{lan12}. 
From now on we set $\varDelta E=\varDelta E^*(c)$ and denote this maximal sensitivity by  $R(c)\equiv R(c,\varDelta E^*(c))$.

\begin{figure}%
 \centering%
 \includegraphics{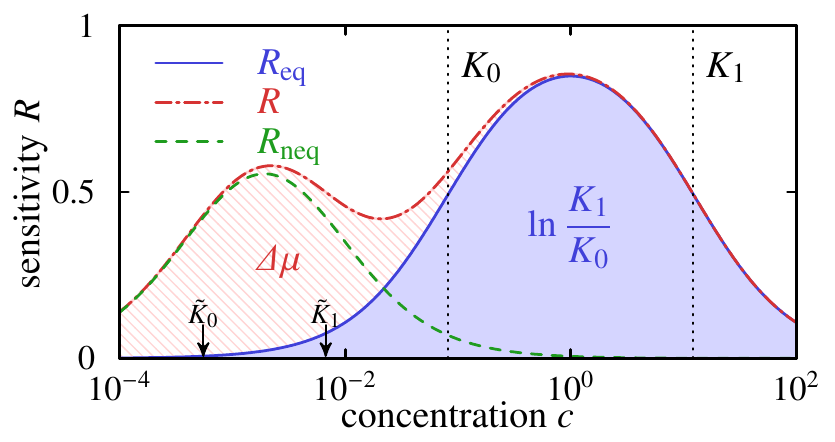}%
 \caption{Increase in sensitivity $R(c)$ through nonequilibrium driving. The full blue region under the curve corresponds to the equilibrium contribution $\ln(K_1/K_0)$, while the 
striped region highlights the additional nonequilibrium enhancement, which is equal to the driving affinity $\varDelta\mu$. The parameters are set to $K_1=1/K_0=\e^{2.5}$, $\alpha=\e^{5}$, and $\varDelta\mu=2.5$.
}%
 \label{fig:area}%
\end{figure}%

For the equilibrium case, expressed in Eq. \eqref{eq:fraction_special_DB}, the sensitivity becomes
\begin{equation}
R_\mathrm{eq}(c)=\frac{c}{c+K_0}-\frac{c}{c+K_1}.
\label{eq:Req}
\end{equation}
Whereas out of equilibrium, with the ratio of probabilities in Eq. \eqref{eq:fraction_special}, we obtain
\begin{equation}
R(c)=R_\mathrm{eq}(c)+R_\mathrm{neq}(c),
\label{eq:R}
\end{equation}
with
\begin{equation}
R_\mathrm{neq}(c)=\frac{c}{c+\big(\frac{K_0}{\alpha}\big)}-\frac{c}{c+\big(\frac{K_0\ee^{\varDelta\mu}}{\alpha}\big)}=\frac{c}{c+\tilde{K_0}}-\frac{c}{c+\tilde{K_1}}.
\label{eq:Rneq}
\end{equation}
Therefore, the effect of adding a driving affinity $\varDelta \mu$ to the single receptor is to increase the sensitivity
by $R_\mathrm{neq}(c)$. Particularly, the sensitivity of this single receptor out of equilibrium is equal to the sensitivity of
a equilibrium model that has a second binding site with dissociation constant $\tilde{K}_a$, as given by \eqref{eq:Keff}. This situation is represented in Fig. \ref{fig:area},
where we show the equilibrium contribution to sensitivity peaking between the concentration range $K_0\le c\le K_1$ and the
nonequilibrium contribution peaking between the range $\tilde{K}_0\le c\le \tilde{K}_1$. Calculating the maximum of $R_\mathrm{neq}(c)$, with $c$ as the optimizing parameter, we obtain the inequality 
\begin{equation}
R_\mathrm{neq}(c)\le \frac{\e^{\varDelta\mu/2}-1}{\e^{\varDelta\mu/2}+1}.
\end{equation} 
We note that an enhancement on sensitivity due to an nonequilibrium driving affinity has been shown in \cite{tu08a} (see also \cite{skog13}).     

As a first main result, we obtain that the integrated sensitivity $I$ has the following simple relation with the driving affinity $\varDelta \mu$,   
\begin{align}
 I&\equiv\int_{-\infty}^{\infty}\dd (\ln c)R(c)
= \varDelta\mu+\ln\frac{K_1}{K_0},
 \label{eq:main_special}
\end{align}
where we used Eqs. \eqref{eq:Req}, \eqref{eq:R} and \eqref{eq:Rneq}. This result for the integrated sensitivity provides a precise
quantification of the effect of free energy dissipation on sensing. The integral represents the area under the curves in Fig. \ref{fig:area}
where the equilibrium contribution $R_\mathrm{eq}(c)$ yields $\ln(K_1/K_0)$ and the nonequilibrium contribution rises the area under the curve by  
$\varDelta \mu$.

From the expressions \eqref{eq:Req} and \eqref{eq:Rneq} it follows that $R_\mathrm{eq}(c)\le 1$ and $R_\mathrm{neq}(c)\le 1$, respectively.
The effect of the driving affinity on the sensitivity is twofold: it can increase the concentration range for which the sensitivity is 
non-negligible and it can increase the sensitivity in the equilibrium range $K_0\le c\le K_1$. The nonequilibrium enhancement 
can even lead to $R(c)>1$ within this region.

\begin{figure}
 \centering
\includegraphics{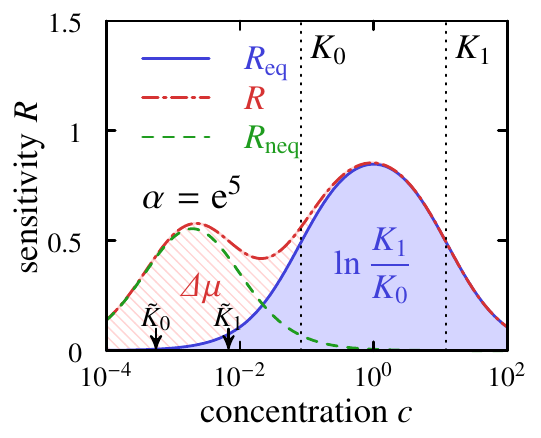}\includegraphics{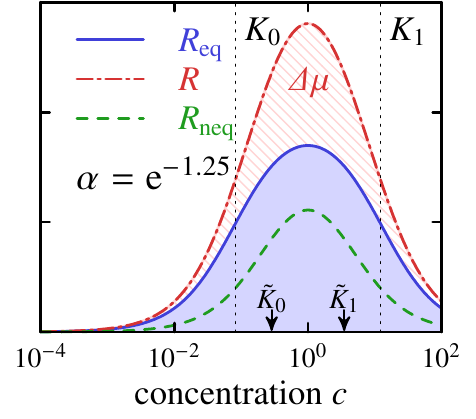}\includegraphics{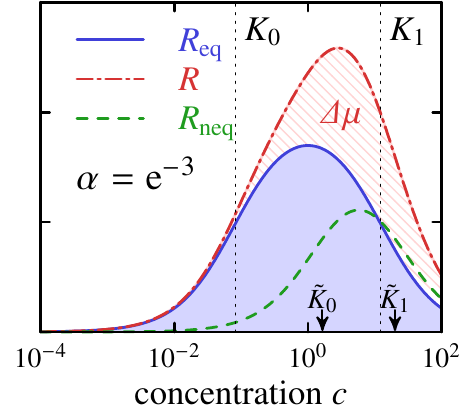}%
\caption{Effect of parameter $\alpha$ on the sensitivity $R(c)$. The parameter $\alpha$ is indicated in the figures, $K_1=1/K_0=\e^{2.5}$, and $\varDelta\mu=2.5$.
For $\alpha= \e^5$ the driving affinity increases the range for which the sensitivity is non-negligible, while in the other two cases we can see a clear 
increase in the sensitivity in the equilibrium range $K_0\le c\le K_1$, with $\alpha=\e^{1.25}$ corresponding to the optimal increase in the equilibrium range. 
}
 \label{fig:3}
\end{figure}

The influence of the parameter $\alpha$ as determined by Eq. \eqref{eq:Rneq} on these two effects
is shown in Fig. \ref{fig:3}, which indicates that there is an optimal $\alpha$ for which the effect of $\varDelta \mu$ is mostly 
to increase the sensitivity in the equilibrium range $K_0\le c\le K_1$. To quantify the enhancement of sensitivity in this equilibrium range we define the
integral 
\begin{align}
 I_{K_0,K_1}^\mathrm{neq}&\equiv\int_{\ln K_0}^{\ln K_1}\dd(\ln c) R_\mathrm{neq}(c)\nonumber\\
 &=\ln\left(\frac{K_1+\frac{K_0}{\alpha}}{K_1+\frac{K_0}{\alpha}\ee^{\varDelta\mu}}\right)-\ln\left(\frac{K_0+\frac{K_0}{\alpha}}{K_0+\frac{K_0}{\alpha}\ee^{\varDelta\mu}}\right),
\label{eq:IK0K1}
\end{align}
We point out that $I_{K_0,K_1}^{\mathrm{neq}}\le \varDelta \mu$ due to Eq. \eqref{eq:main_special} and $I_{K_0,K_1}^{\mathrm{neq}}\le \ln (K_1/K_0)$
due to $R_\mathrm{neq}(c)\le 1$. Maximizing this integral with respect to $\alpha$ we obtain 
\begin{equation}
 I_{K_0,K_1}^{\mathrm{neq},\mathrm{opt}}\equiv
\max_{\alpha}I_{K_0,K_1}^\mathrm{neq}=
\ln\left[\frac{K_1}{K_0}
\left(
\frac{\ee^{\varDelta\mu/2}+\sqrt{\frac{K_0}{K_1}}}{\ee^{\varDelta\mu/2}+\sqrt{\frac{K_1}{K_0}}}
\right)^2
\right],
\label{eq:IK0K1opt}
\end{equation}
where the maximum is obtained for $\alpha=\sqrt{K_0/K_1}\exp(\varDelta\mu/2)$, which leads to dissociation constants that satisfy $\tilde{K}_0\tilde{K}_1=K_0K_1$. 
Hence, expression \eqref{eq:IK0K1opt} provides the optimal sensitivity enhancement due to the driving affinity $\varDelta \mu$ within the equilibrium concentration 
range $K_0\le c\le K_1$ for given $\Delta \mu$, $K_0$, and $K_1$. We note that the effect of increasing the sensitivity beyond the equilibrium range can represent
an important advantage for the cell. This increase is quantified by $I-\ln(K_1/K_0)- I_{K_0,K_1}^\mathrm{neq}$. A more quantitative relation could arise from 
studying the sensitivity integrated over some concentration range of interest. Our choice in Eq. \eqref{eq:IK0K1} is motivated by the fact that $I_{K_0,K_1}^\mathrm{neq}$
is convenient for the analogy between nonequilibrium sensing  and kinetic proofreading.

\section{Analogy with kinetic proofreading}
\label{sec4}

In this section we establish an explicit analogy between sensing with nonequilibrium receptors and kinetic proofreading, with the integrated sensitivity 
in Eq. \eqref{eq:IK0K1} playing the role of the error reduction due to dissipation in kinetic proofreading.

\subsection{Kinetic proofreading}
\begin{figure}%
 \centering%
 \includegraphics{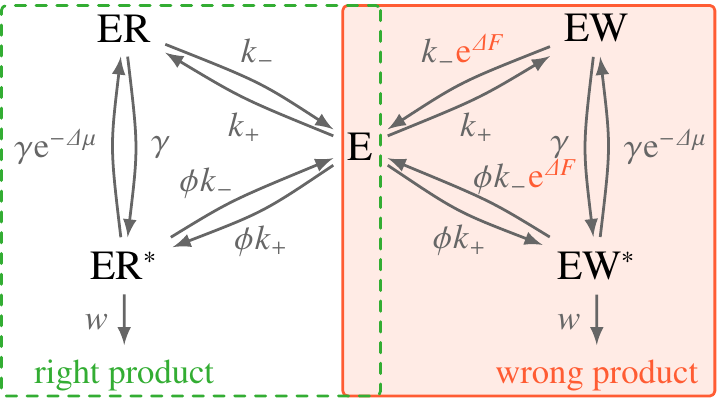}
 \caption{Model for kinetic proofreading. The difference in the transition rates for the two cycles is in the free energy term $\ee^{\varDelta F}$ in 
the transition from EW to E and from EW${}^*$ to E, which is related to the higher free energy of EW in comparison to ER. 
The rate at which information is written $w$ is assumed to be small compared to the other transition rates.}%
 \label{fig:kinetic_proofreading_model}%
\end{figure}%

The model for kinetic proofreading is illustrated in Fig \ref{fig:kinetic_proofreading_model}. Two substrates $\mathrm{S}=\mathrm{R},\mathrm{W}$, with R being the ``right'' substrate and 
W the ``wrong'' substrate, can bind to the enzyme E. In equilibrium, the substrate R is copied to a template with higher probability due to a free energy difference $\varDelta F$.
Specifically, this free energy difference between state EW and ER leads to an equilibrium error
\begin{equation}
\epsilon_\mathrm{eq}= \exp(-\varDelta F),
\end{equation}      
where the error is defined as the ratio between the probability of writing W and the probability of writing R to the template \cite{hopf74,nini75}.  

In the kinetic proofreading scheme phosphorylated forms of the substrates are added, leading to the additional states  $\mathrm{EW}^*$ and $\mathrm{ER}^*$. 
The transitions in Fig. \ref{fig:kinetic_proofreading_model} involve phosphorylation reactions
\begin{equation}
 \mathrm{ES}+\mathrm{ATP}\rightleftharpoons \mathrm{ES}^*+\mathrm{ADP} 
\label{eqreaction1} 
\end{equation}  
and dephosphorylation reactions
\begin{equation}
 \mathrm{ES}^*\rightleftharpoons \mathrm{E}+\mathrm{S}+\mathrm{P_i}. 
\label{eqreaction2} 
\end{equation} 
If a cycle $\mathrm{E}\to \mathrm{ES}\to \mathrm{ES}^*\to \mathrm{E}$ is completed, one ATP is consumed and $\mathrm{ADP}+\mathrm{P_i}$ is produced, leading to a free energy consumption of 
$\varDelta \mu= \mu_\mathrm{ATP}-\mu_\mathrm{ADP}-\mu_\mathrm{P_i}$. The specific transition rates are shown in Fig. \ref{fig:kinetic_proofreading_model},
where $k_-$, $k_+$, $\gamma$, and $\phi$ are kinetic parameters. Moreover, $w$ is the rate at which the substrate S is written to the template, which we assume to be much slower than the other
transition rates, i.e., we assume the limit $w\to 0$.

The error is given by 
\begin{align}
 \epsilon\equiv \frac{P_{\mathrm{EW}^*}}{P_{\mathrm{ER}^*}}={}&
 \left(
 \frac{
(\ee^{-\varDelta\mu}+\phi)\gamma+\phi k_-
}{
(\ee^{-\varDelta\mu}+\phi)\gamma\ee^{\varDelta F}+\phi k_-\ee^{2\varDelta F}
}
\right)
\left(\frac{(1+\phi)\gamma+\phi k_-\ee^{\varDelta F}}{(1+\phi)\gamma+\phi k_-}\right),
\label{eq:error_rate}
\end{align}
where $P_{\mathrm{EW}^*}$ and $P_{\mathrm{ER}^*}$ denote the stationary probabilities of states $\mathrm{EW}^*$ and $\mathrm{ER}^*$, respectively.
Hence, as first observed by Hopfield and Ninio \cite{hopf74,nini75}, with energy dissipation the error can be smaller than $\epsilon_\mathrm{eq}$. The maximal error
reduction $\epsilon/\epsilon_\mathrm{eq}= \ee^{-\varDelta F}$ takes place for an appropriate choice of the kinetic parameters and the formal limit $\varDelta \mu\to \infty$.

\subsection{Non-equilibrium sensing vs. kinetic proofreading}

\begin{figure}%
 \centering%
\subfloat[][]{%
\includegraphics{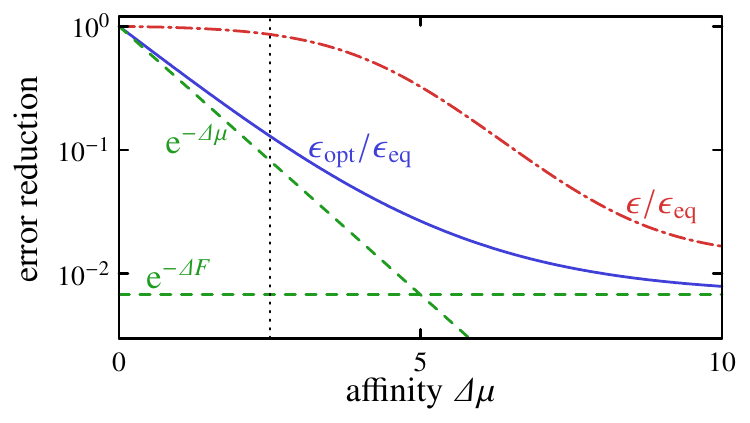}%
\label{fig:error_dEa}%
}%
\subfloat[][]{%
\includegraphics{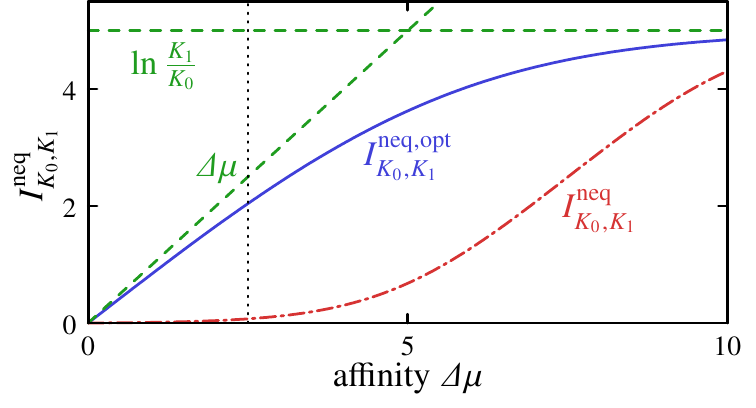}%
\label{fig:error_dEb}%
}

\subfloat[][]{%
\includegraphics{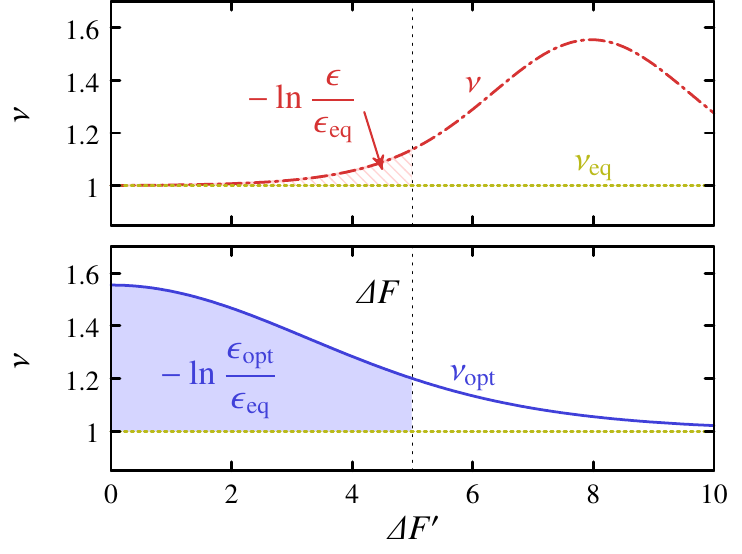}%
\label{fig:error_dEc}%
}
\subfloat[][]{%
\includegraphics{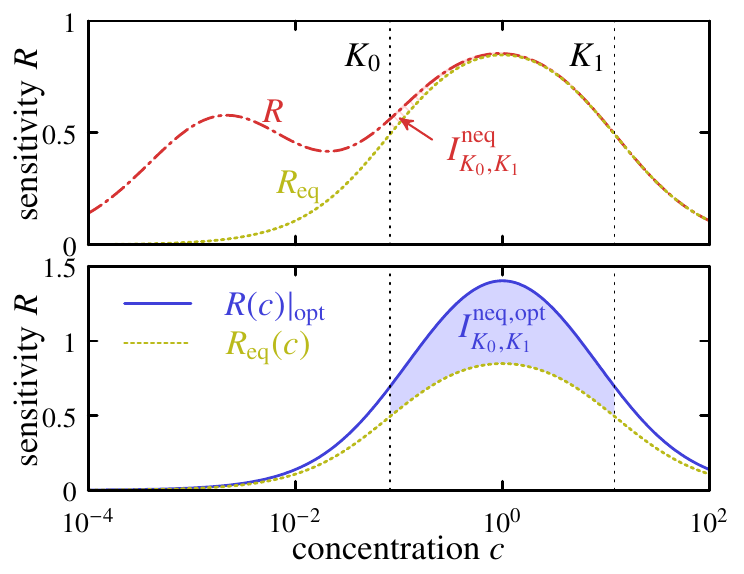}%
\label{fig:error_dEd}%
}
 \caption{Kinetic proofreading (left panel) versus nonequilibrium sensing (right panel).
 \textbf{ (a)} Error reduction $\epsilon/\epsilon_\mathrm{eq}$ and optimal error reduction $\epsilon_\mathrm{opt}/\epsilon_\mathrm{eq}$, as given by \eqref{eq:error_min}, as functions of the 
affinity $\varDelta\mu$. The green dashed lines indicate the asymptotically reached bounds $\epsilon_\mathrm{opt}/\epsilon_\mathrm{eq}\ge\ee^{-\varDelta \mu}$ and $\epsilon_\mathrm{opt}/\epsilon_\mathrm{eq}\ge\ee^{-\varDelta F}$.
\textbf{(b)} Nonequilibrium contribution to the sensitivity integrated in the equilibrium range 
$I_{K_0,K_1}^\mathrm{neq}$ and its optimized value $I_{K_0,K_1}^{\mathrm{neq},\mathrm{opt}}$, as given by \eqref{eq:IK0K1opt}, 
as functions of the affinity $\varDelta\mu$. The green dashed lines indicate the asymptotically reached bounds $I_{K_0,K_1}^\mathrm{neq}\le\varDelta \mu$ and $I_{K_0,K_1}^\mathrm{neq}\le\ln(K_1/K_0)$. 
\textbf{(c)} The discriminatory index $\nu$ as a function of the free energy difference $\varDelta F'$. The lower panel shows
 $\nu_\mathrm{opt}\equiv-\partial_{\varDelta F}\ln \epsilon_\mathrm{opt}(\varDelta F,\varDelta\mu)$, i.e., the discriminatory index associated with the minimal error \eqref{eq:error_min}. 
The highlighted areas illustrate  relation \eqref{eq:Inu-1}. \textbf{(d)} The sensitivity $R(c)$ and the sensitivity $R(c)|_\mathrm{opt}$, which is associated with $I_{K_0,K_1}^{\mathrm{neq},\mathrm{opt}}$. The highlighted areas illustrate 
relation \eqref{eq:IK0K1}. Parameters are set in the following way: $\varDelta F=\ln(K_1/K_0)=5$ with $K_1=1/K_0$ in (a) and (b); $\varDelta\mu=2.5$ in (c) and (d); in (b) and (d) $I_{K_0,K_1}^\mathrm{neq}$ is obtained 
from \eqref{eq:IK0K1} with $\alpha=\ee^5$; in (a) and (c) $\epsilon/\epsilon_\mathrm{eq}$ is obtained from \eqref{eq:error_rate} with $k_-=\gamma=1$, $\phi=10^{-4}$. The dotted vertical line in (a) and (b)
indicate the affinity $\varDelta\mu=2.5$. The dotted vertical line in (c) indicates $\varDelta F=5$.}
 \label{fig:error_dE}%
\end{figure}%

The minimal error $\epsilon_\mathrm{opt}$ for fixed free energy difference $\varDelta F$ and driving affinity $\varDelta\mu$, that is
obtained by optimizing $\epsilon$ in Eq. \ref{eq:error_rate} with respect to the kinetic parameters, is given by \cite{qian06,qian07}
\begin{equation}
 \frac{\epsilon_\mathrm{opt}}{\epsilon_\mathrm{eq}}(\varDelta F,\varDelta\mu)=\ee^{-\varDelta F}\left(\frac{\ee^{\frac{\varDelta\mu}{2}}+\ee^{\frac{\varDelta F}{2}}}{\ee^{\frac{\varDelta\mu}{2}}+\ee^{-\frac{\varDelta F}{2}}}\right)^2,
\label{eq:error_min}
\end{equation}
Since this function  is bounded by $\ee^{-\varDelta \mu}$ and by $\ee^{-\varDelta F}$, the following inequality holds \cite{qian06},
\begin{equation}
\epsilon\ge\epsilon_\mathrm{opt}\ge\exp(-\varDelta F-\varDelta\mu).
\label{eq:ineerr}
\end{equation}

Comparing expression \eqref{eq:error_min} for the maximal error reduction in kinetic proofreading with expression \eqref{eq:IK0K1opt} for the maximal increase in the integrated sensitivity in nonequilibrium sensing, a quite
transparent analogy arises, as shown in Figs. \ref{fig:error_dEa} and \ref{fig:error_dEb}. Both expressions are the same with the increase in sensitivity in the equilibrium range $I^{\mathrm{neq},\mathrm{opt}}_{K_0,K_1}$ being analogous 
to $-\ln(\epsilon_\mathrm{opt}/\epsilon_\mathrm{eq})$ and the ratio of the dissociation constants $K_1/K_0$ being analogous to $\ee^{\varDelta F}$. Whereas in kinetic proofreading a driving affinity $\varDelta\mu$ decreases the error, 
in nonequilibrium sensing $\varDelta\mu$ increases the integrated sensitivity in the equilibrium range $K_0\le c\le K_1$.
  
A recently introduced quantity in kinetic proofreading is the discriminatory index \cite{muru14}
\begin{equation}
 \nu(\varDelta F)\equiv-\frac{\partial }{\partial \varDelta F}\ln\epsilon,
 \label{eq:nu_def}
\end{equation}
where $\epsilon$ is given by \eqref{eq:error_rate}. Due to $\epsilon_\mathrm{eq}=\ee^{-\varDelta F}$ the discriminatory index is $\nu_\mathrm{eq}=1$ in equilibrium, with a larger index $\nu\ge1$ requiring energy dissipation.
We can rewrite \eqref{eq:nu_def} as   
\begin{equation}
-\ln\frac{\epsilon}{\epsilon_\mathrm{eq}}=\int_0^{\varDelta F}\dd \varDelta F'[\nu(\varDelta F')-1].
\label{eq:Inu-1}
\end{equation}
Comparing this equation with \eqref{eq:IK0K1} we observe that the discriminatory index is analogous to the sensitivity $R(c)$, with
$\nu(\varDelta F')-1$ being the nonequilibrium contribution. In Figs. \ref{fig:error_dEc} and \ref{fig:error_dEd} we show the comparison between
discriminatory index in proofreading and sensitivity in nonequilibrium sensing. Murugan et al. \cite{muru14} have shown that the integral from $-\infty$ to $\infty$ of $\nu(\varDelta F)-1$
can be equal to $\varDelta \mu$. This result is equivalent to our equality \eqref{eq:main_special}.

\section{Effect of the occupancy of the receptor on phosphorylation and dephosphorylation rates}
\label{sec:multiple_links}

\begin{figure}%
\centering
\subfloat[][]{%
\includegraphics{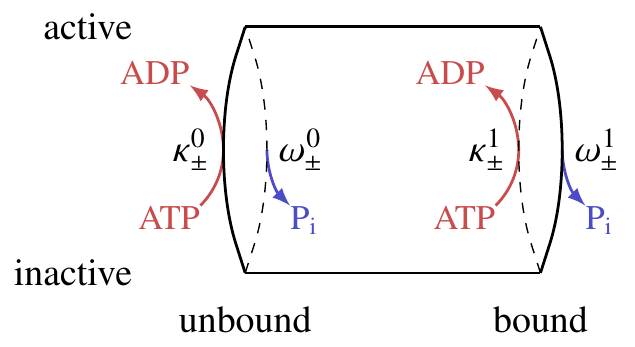}
\label{fig:multiple_linksA}%
}\qquad
\subfloat[][]{%
\includegraphics{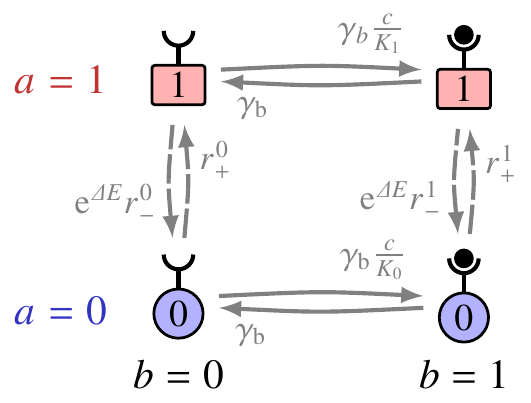}
 \label{fig:multiple_linksB}%
}
\caption{Four-state system with phosphorylation and dephosphorylation for both $b=0$ and $b=1$. \textbf{(a)} Full model with two links for the vertical transitions. The dashed links indicate transition rates that are zero 
in the model of Fig. \ref{fig:general4state}. \textbf{(b)} Total rates as given by \eqref{eq:A_rpm} and \eqref{eq:A_rpm2}. The coarse-grained entropy production \eqref{eqcoarse} is calculated with these total transition rates.}%
\label{fig:multiple_links}%
\end{figure}
We now generalize the model from Fig. \ref{fig:general4state} to include phosphorylation and dephosphorylation reactions for both $b=0$ and $b=1$. With this generalization
there are two links for the vertical transitions in Fig. \ref{fig:multiple_linksA}. These reactions happen with transition rates
\begin{equation}
 \raisebox{-0.25\height}{\includegraphics{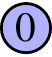}}+ \mathrm{ATP}
  \xrightleftharpoons[{\kappa_-^b}]{\kappa_+^b}
\raisebox{-0.25\height}{\includegraphics{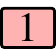}}
  +\mathrm{ADP}
  \xrightleftharpoons[{\omega_-^b}]{\omega_+^b}
	\raisebox{-0.25\height}{\includegraphics{Figures/symbol5}}
	 +\mathrm{ADP}+\mathrm{P_i},
\label{eq:kappa}
\end{equation}
where $b=0,1$. For thermodynamic consistency, the following constraints must be fulfilled:
\begin{equation}
\varDelta\mu=\ln\frac{\kappa_+^b\omega_+^b}{\kappa_-^b\omega_-^b}
\label{eq:A_wk-cycle}
\end{equation}
for $b=0,1$,
\begin{equation}
\ln\frac{\kappa^0_+\omega^1_+}{\kappa^0_-\omega^1_-}= \varDelta\mu+\ln\frac{K_1}{K_0},
\label{eq:A_wk-cycle2}
\end{equation}
and
\begin{equation}
\ln\frac{\kappa^1_+\omega^0_+}{\kappa^1_-\omega^0_-}= \varDelta\mu-\ln\frac{K_1}{K_0},
\label{eq:A_wk-cycle3}
\end{equation}
where we used the free energy \eqref{eq:Fab} for the second and third equations. Whereas the 
presence of two links is important for calculating the rate of dissipation in this model \cite{seif12},
for the purpose of calculating the stationary probabilities we consider the total transition rates from inactive to active
\begin{equation}
 r_{+}^{b}\equiv\kappa_+^b+\omega_-^b
\label{eq:A_rpm}
\end{equation}
and from active to inactive
\begin{equation}
 \e^{\varDelta E}r_{-}^{b}\equiv\kappa_-^b+\omega_+^b,
\label{eq:A_rpm2}
\end{equation}
which are indicated in Fig. \ref{fig:multiple_linksB}. We choose the rates $\kappa_-^b$ and $\omega_+^b$ to be proportional to $\e^{\varDelta E}$, which leads to 
$r_{\pm}^{b}$ independent of $\varDelta E$. 
Assuming that the binding/unbinding transitions are much faster, the ratio of stationary probabilities \eqref{eq:fraction_special} for
this more general model becomes   
\begin{align}
\frac{P_0}{P_1}&=\e^{\varDelta E}\left(\frac{1+\frac{c}{K_0}}{1+\frac{c}{K_1}}\right)\left(\frac{r_-^0+(\frac{c}{K_1})r_-^1}{r_+^0+(\frac{c}{K_0})r_+^1}\right)
\equiv \e^{\varDelta E}\left(\frac{1+\frac{c}{K_0}}{1+\frac{c}{K_1}}\right)\chi(c).
\label{eq:A_P0P1_chi}
\end{align}
As in section \ref{sec3}, the sensitivity \eqref{eq:fmax_def} is maximized for $P_0/P_1=1$, which is achieved at 
\begin{equation}
\varDelta E^*(c)=\ln\left(\frac{1+\frac{c}{K_1}}{1+\frac{c}{K_0}}\right)-\ln \chi(c),
\label{eq:E*2}
\end{equation}
The equilibrium contribution to the maximal sensitivity is given by \eqref{eq:Req}, while the nonequilibrium contribution is 
\begin{equation}
 R_\mathrm{neq}(c)\equiv\frac{\del}{\del(\ln c)}\ln\chi(c)=\frac{1}{c}\frac{\chi'(c)}{\chi(c)}.
 \label{eq:A_Rneq_chi}
\end{equation}
The integrated sensitivity then becomes
\begin{align}
 I&=\int_{-\infty}^{\infty}\dd(\ln c)[ R_\mathrm{neq}(c)+ R_\mathrm{eq}(c)]=\ln\frac{K_1}{K_0}+\ln\frac{\chi(\infty)}{\chi(0)}\nonumber\\
 &=\ln\left(\frac{r_+^0r_-^1}{r_-^0r_+^1}\right).
 \label{eq:A_I}
\end{align}
For $\varDelta\mu\ge0$, from Eqs. \eqref{eq:A_wk-cycle}, \eqref{eq:A_rpm}, and \eqref{eq:A_rpm2}, we obtain
\begin{equation}
\frac{\omega_-^b}{\omega_+^b}\le\frac{r_+^b}{\e^{\varDelta E}r_-^b}\le\frac{\kappa_+^b}{\kappa_-^b}.
\label{eq:A_ineq_chain}
\end{equation}
With these inequalities, we obtain that the integrated sensitivity \eqref{eq:A_I} is bounded by   
\begin{equation}
 I=\ln\left(\frac{r_+^0r_-^1}{r_-^0r_+^1}\right)\le\ln\frac{\kappa^0_+\omega^1_+}{\kappa^0_-\omega^1_-}=\ln\frac{K_1}{K_0}+\varDelta \mu,
 \label{eq:main_special2}
\end{equation}
where we used \eqref{eq:A_wk-cycle2} in the last equality. As shown in \ref{sec:Nsites}, this inequality can also be generalized to an
arbitrary number of binding sites.

The influence of how the occupancy of the receptor affects the reaction rates for activity on the relation between 
the integrated sensitivity $I$ and the driving affinity $\varDelta \mu$ can be seen with the following examples.
First, if we choose transition rates satisfying the relation
\begin{equation}
\frac{K_0r_-^1}{K_1r_+^1}=\frac{r_-^0}{r_+^0},
\end{equation}
the function $\chi(c)$ in \eqref{eq:A_P0P1_chi} becomes independent of $c$. In this case, from \eqref{eq:A_Rneq_chi} we obtain $R_\mathrm{neq}(c)=0$,
which implies $I= \ln (K_1/K_0)$. Hence, it is possible to have a dissipative model with ATP consumption that has the same sensitivity as the equilibrium case.
Second, we consider the case where phosphorylation happens only if the receptor is bound and dephosphorylation occurs only if the receptor is unbound, which is
the opposite of the model in Fig. \ref{fig:general4state}. In this case $\kappa_\pm^0=\omega_\pm^1=0$, leading to 
\begin{equation}
 I=\ln\frac{\kappa^1_-\omega^0_-}{\kappa^1_+\omega^0_+}=\ln\frac{K_1}{K_0}-\varDelta \mu,
 \label{eq:A_I_ineq_eq}
\end{equation}     
where we used \eqref{eq:main_special2} and \eqref{eq:A_wk-cycle3}. This result shows that the integrated sensitivity can also decrease with energy dissipation.
The regime for which the integrated sensitivity is decreased by $\varDelta \mu$ is equivalent to an anti-proofreading regime recently studied in \cite{muru14}.

A more precise analysis of the relation between $I$ and energy dissipation can be achieved by considering the entropy production $\sigma$ \cite{seif12}. For the  
present model this entropy production is the rate of ATP consumption. Using the stationary probability $P_{a,b}$ we define the probability current
$J_b\equiv P_{0,b}\kappa_+^b-P_{1,b}\kappa_-^b$. With this current the entropy production can be written as
\begin{equation}
\sigma= (J_0+J_1) \varDelta \mu,
\end{equation}  
by using the fact that $\sigma$ is a sum of currents multiplying cycle affinities \cite{seif12}. The energy dissipation $\sigma$ is
non-zero whenever $\varDelta \mu\neq 0$. Besides $\sigma$, we can consider a coarse-grained entropy production $\tilde{\sigma}$, which does not take into account the
two channels for the vertical transitions in Fig. \ref{fig:multiple_linksA}: it is obtained by considering the single links with rates $r_{\pm}^b$ in Fig. \ref{fig:multiple_linksB}. This coarse-grained entropy 
production provides a lower bound on the full entropy production, i.e.,
$\sigma\ge \tilde{\sigma}$ \cite{espo12}. For the model in Fig. \ref{fig:multiple_linksB},
\begin{equation}
\tilde{\sigma}= J\mathcal{A}.
\label{eqcoarse}
\end{equation}  
where $J\equiv r_+^0P_{0,0}-\e^{\varDelta E}r_-^0P_{1,0}$ and 
\begin{equation}
\mathcal{A}\equiv \ln\left(\frac{r_+^0r_-^1}{r_-^0r_+^1}\right)-\ln \frac{K_1}{K_0}
\end{equation}
is an effective affinity. From relation \eqref{eq:main_special2} we obtain 
\begin{equation}
\mathcal{A}= I-\ln \frac{K_1}{K_0},
\end{equation}
which is the nonequilibrium contribution to the integrated sensitivity $I$. The effective affinity associated with the coarse-grained entropy production
determines three different regimes for nonequilibrium sensing. For $\mathcal{A}>0$ the integrated sensitivity is increased in relation to its equilibrium value. 
If $\mathcal{A}=0$, which implies $\tilde{\sigma}=0$, the energy dissipation has 
no effect on sensitivity. If $\mathcal{A}<0$ then the inequality $\tilde{\sigma}\ge0$ implies $J<0$. In this last regime energy dissipation decreases the integrated sensitivity.

\section{Conclusion}\label{sec:conclusions}

We have characterized the enhancement of sensitivity by a nonequilibrium driving affinity that arises from ATP hydrolysis in the chemical reactions involving an activity change.
For the single receptor model from Sec. \ref{sec:Main_result}, the integrated sensitivity $I$ was shown to have a simple relation with the driving affinity in Eq. \eqref{eq:main_special}.
We have shown that a dissipative sensing model can lead to both an increase in the concentration range for which the sensitivity is non-negligible and an increase in the sensitivity
in the equilibrium range. The second effect is quantified  by $I_{K_0,K_1}^\mathrm{neq}$, which is defined in Eq. \eqref{eq:IK0K1}.

We have shown that nonequilibrium sensing is equivalent to kinetic proofreading, with the analogous parameters, observables and relations summarized in Tab. \ref{tab:analogies}. Most prominently, while in nonequilibrium sensing 
a driving affinity leads to an increase in the sensitivity integrated over the equilibrium range, in kinetic proofreading a driving affinity decreases the error. In kinetic proofreading the equivalent of sensitivity is
the discriminatory index introduced in \cite{muru14}.

\begin{table}%
 \centering%
\begin{tabular}{c|c|c|c|c}
Nonequilibrium sensing     &$I_{K_0K_1}^\mathrm{neq}$ &$R$&$\ln (K_1/K_0)$& $I\le\ln(K_1/K_0)+\varDelta\mu$\\\hline
Kinetic proofreading& $-\ln(\epsilon/\epsilon_\mathrm{eq})$&$\nu$&$\varDelta F$&$\epsilon\ge\exp(-\varDelta F-\varDelta\mu)$
 \end{tabular}%
\caption{Nonequilibrium sensing compared to kinetic proofreading.}%
\label{tab:analogies}%
\end{table}%

The influence of the occupancy of the receptor on the phosphorylation and dephosphorylation rates is of fundamental importance for the relation between integrated sensitivity and the affinity. As we have shown in
section \ref{sec:multiple_links}, it is even possible to have a regime where energy dissipation leads to a decrease on the integrated sensitivity, which is analogous to the anti-proofreading regime from \cite{muru14}.     

Our results demonstrate that measurements of the integrated sensitivity could unveil how the occupancy of the receptor affects the 
phosphorylation and dephosphorylation rates. It is certainly intriguing to speculate whether real chemotaxis networks evolved in such a
way that this influence optimizes the enhancement of sensitivity due to energy consumption.

\ack
D. H. acknowledges helpful discussions with J. M. S\'anchez.

\appendix

\section{Generalization to an arbitrary number of binding sites}
\label{sec:Nsites}

\begin{figure}
\centering
  \includegraphics{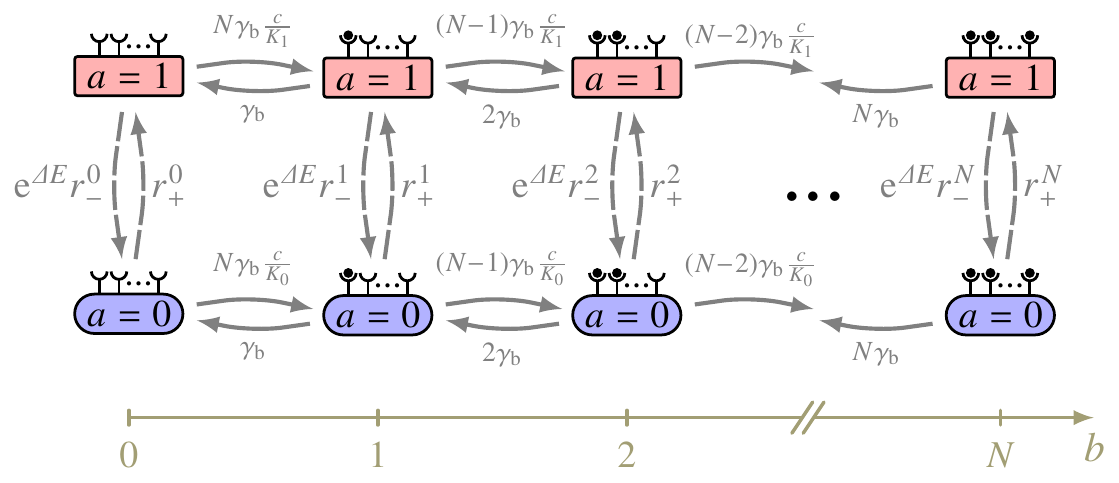}
\caption{Generalization of the single receptor model in the main text to an arbitrary number of binding sites $N$. The rates $r_+^b$ and $\ee^{\varDelta E} r_-^b$ are 
defined in \eqref{eq:A_rpm} and \eqref{eq:A_rpm2}, respectively.}
\label{fig:nu_bindingsites}%
\end{figure}

In this Appendix, we consider a generalization of the model studied in the main text for an arbitrary number of binding sites. In this case, the variable $b$ takes the values
$b=0,1,\ldots,N$, where $N$ the number of binding sites. The transition rates for this more general model are shown in Fig. \ref{fig:nu_bindingsites}. Rates involving
a change in activity  are given by \eqref{eq:A_rpm} and \eqref{eq:A_rpm2}. Rates related to a change in the occupancy of receptor must fulfill the generalized detailed balance relation
with respect to the free energy \eqref{eq:Fab}. We set these rates as follows. The binding rate from $b$ to $b+1$ is $w_{b,b+1}^{a}= \gamma_b(N-b)c/K_a$, where the factor $N-b$ comes from the fact that there are
$N-b$ free receptors for the ligand to bind; the unbinding rate from $b$ to $b-1$ is $w_{b,b-1}^{a}= \gamma_bb$, where the factor $b$ is related to the number of bound receptors.
 
We assume that the binding events are much faster than changes in activity. In this case, the stationary conditional probability reads 
\begin{equation}
P(b|a)\equiv\frac{P_{a,b}}{P_a}=\binom{N}{b}\left(\frac{c}{K_a}\right)^b\bigg/\left(1+\frac{c}{K_a}\right)^{N},
\label{eq:binomialN}
\end{equation}
leading to
\begin{equation}
\frac{P_0}{P_1}=\e^{\varDelta E}\left(\frac{1+\frac{c}{K_0}}{1+\frac{c}{K_1}}\right)^N\chi(c),
\label{eq:A_P0P1_chiN}
\end{equation}
where
\begin{equation}
\chi(c)=\frac{\sum_{b=0}^N\binom{N}{b}\big(\frac{c}{K_1}\big)^br_-^b}{\sum_{b=0}^N\binom{N}{b}\big(\frac{c}{K_0}\big)^br_+^b}.
\label{eq:A_chiN}
\end{equation}
This expression generalizes \eqref{eq:A_P0P1_chi} to the case of $N$ binding sites. Following the same procedure from section \ref{sec:multiple_links}, similarly to \eqref{eq:E*2}  
the sensitivity \eqref{eq:fmax_def} is maximized for
\begin{equation}
\varDelta E^*(c)=N\ln\left(\frac{1+\frac{c}{K_1}}{1+\frac{c}{K_0}}\right)-\ln \chi(c),
\label{eq:E*3}
\end{equation}
where, similarly to \eqref{eq:A_Rneq_chi}, the nonequilibrium contribution to sensitivity becomes
\begin{equation}
 R_\mathrm{neq}(c)=\frac{\del}{\del(\ln c)}\ln\chi(c)=\frac{1}{c}\frac{\chi'(c)}{\chi(c)}.
 \label{eq:A_Rneq_chi2}
\end{equation}
Hence, the integrated sensitivity reads
\begin{align}
 I & =\int_{-\infty}^\infty\dd(\ln c)\big[ R_\mathrm{eq}(c)+R_\mathrm{neq}(c)\big]=N\ln\frac{K_1}{K_0}+\ln\frac{\chi(\infty)}{\chi(0)}\nonumber\\
 & =\ln\left(\frac{r_+^0}{r_-^0}\frac{r_-^N}{r_+^N}\right).
\end{align}
The transition rates for this model must fulfill the constraint
\begin{equation}
\ln\left(\frac{\kappa_+^0}{\kappa_-^0}\frac{\omega_+^N}{\omega_-^N}\right)=N\ln\frac{K_1}{K_0}+\varDelta\mu.
\end{equation}
From the inequalities \eqref{eq:A_ineq_chain} we finally obtain
\begin{equation}
I\le N\ln\frac{K_1}{K_0}+\varDelta\mu,
\label{eq:A_I_ineq_N}
\end{equation}
which generalizes inequality \eqref{eq:main_special2} to the case of $N$ binding sites.


\section*{References}

\bibliography{Sensitivity}

 \end{document}